\documentclass[usenatbib,letters]{mnras}
\usepackage{pdflscape}
\usepackage{color}
\usepackage{IEEEtrantools} %
\usepackage{amsmath} %
\usepackage{amssymb} %
\usepackage{bm} %
\usepackage{upgreek} %
\usepackage[pdftex]{graphicx} %
\usepackage{multirow}
\usepackage{textcomp}
\usepackage{CJK}
\usepackage{rotating}
\usepackage{txfonts}
\usepackage{footnote}
\usepackage[dvipsnames]{xcolor}

\def\araa{ARA\&A}
\def\apj{ApJ}
\def\apjl{ApJ}
\def\apjs{ApJS}

\def\aap{A\&A}

\def\mnras{MNRAS}

\def\nat{Nature}

\def\physrep{Physics Reports}

\newcommand{\fig}[1]{\text{Fig.~\ref{#1}}}

\newcommand{\teff}{T_{\mathrm{eff}}}
\newcommand{\lgg}{\log{g}}

\newcommand{\vsini}{\varv\sin\iota}

\newcommand{\feh}{\mathrm{\left[Fe/H\right]}}

\newcommand{\dex}{\mathrm{dex}}

\newcommand{\halpha}{\mathrm{H\upalpha}}
\newcommand{\hbeta}{\mathrm{H\upbeta}}
\title[Lithium abundances measured in warm dwarf stars]{The GALAH Survey: A new constraint on cosmological lithium and Galactic lithium evolution from warm dwarf stars} 
\author[X.~D.~Gao et al.]{Xudong~Gao$^{1,2}$\thanks{E-mail:gao@mpia.de},
Karin~Lind$^{1,3}$, Anish~M.~Amarsi$^{1,4}$, 
Sven~Buder$^{5,6,1,2}$, 
Joss~Bland-Hawthorn$^{7,6}$, \newauthor
Simon~W.~Campbell$^{8}$,
Martin~Asplund$^{5,6}$, Andrew~R.~Casey$^{9,8}$, 
Gayandhi~M.~De~Silva$^{10,6}$,  \newauthor 
Ken~C.~Freeman$^{5}$, 
Michael~R.~Hayden$^{7,6}$, 
Geraint~F.~Lewis$^{7}$, Sarah~L.~Martell$^{11,6}$,  \newauthor
Jeffrey~D.~Simpson$^{11}$, 
Sanjib~Sharma$^{7,6}$,
Daniel~B.~Zucker$^{12,13,6}$,
Toma\v{z}~Zwitter$^{14}$,  \newauthor
Jonathan~Horner$^{15}$, 
Ulisse~Munari$^{16}$, 
Thomas~Nordlander$^{5,6}$, Dennis~Stello$^{11,6,17}$, \newauthor
Yuan-Sen~Ting$^{18,19,20,5}$, Gregor~Traven$^{21}$, 
Robert~A.~Wittenmyer$^{22}$ and~the~GALAH~collaboration
\newauthor \\
$^{1}$ Max-Planck-Institut f\"ur Astronomie (MPIA), K\"onigstuhl 17, D-69117 Heidelberg, Germany \\
$^{2}$Fellow of the International Max Planck Research School for Astronomy \& Cosmic Physics at the University of Heidelberg, Germany\\
$^{3}$Department of Astronomy, Stockholm University, AlbaNova, Roslagstullbacken 21, SE-10691 Stockholm, Sweden\\
$^{4}$Theoretical Astrophysics, Department of Physics and Astronomy, Uppsala University, Box 516, SE-751 20 Uppsala, Sweden\\
$^{5}$Research School of Astronomy \& Astrophysics, Australian National University, ACT 2611, Australia\\
$^{6}$Center of Excellence for All-Sky Astrophysics in Three Dimensions (ASTRO-3D), Australia\\
$^{7}$Sydney Institute for Astronomy (SIfA), School of Physics, A28, The University of Sydney, NSW 2006, Australia\\
$^{8}$School of Physics and Astronomy, Monash University, Clayton VIC 3800, Australia\\
$^{9}$Monash Centre for Astrophysics, Monash University, Clayton VIC 3800, Australia \\
$^{10}$Australian Astronomical Optics, Macquarie University, 105 Delhi Rd, North Ryde, NSW 2113, Australia \\
$^{11}$School of Physics, University of New South Wales, Sydney, NSW 2052, Australia\\
$^{12}$Department of Physics and Astronomy, Macquarie University, Sydney, NSW 2109, Australia \\
$^{13}$Macquarie University Research Centre for Astronomy, Astrophysics \& Astrophotonics, Sydney, NSW 2109, Australia\\
$^{14}$Faculty of Mathematics and Physics, University of Ljubljana, Jadranska 19, 1000 Ljubljana, Slovenia\\
$^{15}$University of Southern Queensland, Toowoomba, Queensland 4350, Australia\\
$^{16}$INAF Astronomical Observatory of Padova, 36012 Asiago, Italy \\
$^{17}$Stellar Astrophysics Centre, Department of Physics and Astronomy, Aarhus University, DK-8000, Aarhus C, Denmark\\
$^{18}$Institute for Advanced Study, Princeton, NJ 08540, USA \\
$^{19}$Department of Astrophysical Sciences, Princeton University, Princeton, NJ 08544, USA \\
$^{20}$Observatories of the Carnegie Institution of Washington, 813 Santa Barbara Street, Pasadena, CA 91101, USA \\
$^{21}$Lund Observatory, Department of Astronomy and Theoretical Physics, Box 43, SE-221 00 Lund, Sweden \\
$^{22}$University of Southern Queensland, Computational Engineering and Science Research Centre, Toowoomba, Queensland 4350, Australia\\}

\begin{document}
\begin{CJK*}{UTF8}{gbsn}
\date{Accepted ---. Received ---; in original form ---}
\pagerange{\pageref{firstpage}--\pageref{lastpage}} \pubyear{---}

\maketitle 
\end{CJK*}
\label{firstpage}
\begin{abstract}
    Lithium depletion and enrichment in the cosmos
    is not yet well understood.  To help tighten 
    constraints on stellar and Galactic
    evolution models,
    we present the largest high-resolution analysis of 
    Li abundances A(Li) to date,
    with results for over $100\,000$ GALAH field stars spanning effective
    temperatures $5900\,\mathrm{K} \lesssim\teff \lesssim7000\,\mathrm{K}$ and metallicities $-3 \lesssim \feh \lesssim +0.5$.  
    We separated these stars into two groups,
    on the warm and cool side of the so-called Li-dip,
    a localised region of the Kiel diagram
    wherein lithium is severely depleted.
    We discovered that stars in these two groups
    show similar trends in the A(Li)-$\feh$ plane, but 
    with a roughly constant offset in A(Li) of $0.4\,\dex$, the warm group having higher Li abundances.  
    At $\rm[Fe/H]\gtrsim-0.5$, a significant increasing in Li abundance with increasing metallicity is evident in both groups, 
    signalling the onset of significant Galactic production.
    At lower metallicity, stars in the cool group sit on the Spite plateau, showing a reduced lithium of around $0.4\,\dex$ relative to the 
    primordial value predicted from Big Bang nucleosynthesis (BBN).
    However, stars in the warm group between $\feh = -1.0$ and $-0.5$, form an elevated plateau that is
    largely consistent with the BBN prediction. This may indicate that these stars in fact preserve the primordial Li produced in the early Universe.
\end{abstract}

\section{Introduction}
\label{sect:introduction}

Lithium is a fragile element that can be destroyed by proton capture
reactions at relatively low temperatures ($\sim\!2.5\times10^6\,\textrm{K}$) in
stellar interiors \citep{1997ARA&amp;A..35..557P}.  Standard stellar evolution
models suggest that the convective envelopes are weakly developed in low-mass
unevolved (main-sequence) stars with effective temperatures ($\teff$) larger
than $6000\,\mathrm{K}$, thus precluding the surface Li from reaching the
interior to be destroyed \citep{1990ApJS...73...21D}. As such, the unevolved,
metal-poor stars are expected to retain near-primordial Li abundances,
providing an opportunity to put constraints on Big Bang nucleosynthesis (BBN).
However, a significant difference has been found between the Li abundance
measured from very metal-poor stars in the Galaxy 
that fall on the so-called ``Spite plateau'' at A(Li)\footnote{A(Li) =
$\log{\frac{n_{\mathrm{Li}}}{n_{\mathrm{H}}}}+12$, where $n_{\mathrm{Li}}$ and
$n_{\mathrm{H}}$ are the the number densities of lithium and hydrogen,
respectively.} $\approx2.2$ \citep{1982A&amp;A...115..357S}, and the prediction
by standard BBN models
A(Li) $=2.75\pm0.02$
\citep{2018PhR...754....1P}.
This is the well-known Cosmological Lithium
Problem \citep{2012MSAIS..22....9S}.
It has been suggested that the metal-poor stars we observe have undergone lithium depletion by an amount that may be quantified by comparing stellar abundances to evolutionary models with atomic diffusion and an unknown source of mixing in the stellar interior \citep{2005ApJ...619..538R,2006Natur.442..657K}. The results do alleviate the tension to standard Big Bang, but may do not fully bridge the gap as initial abundances of order $\sim\!2.57$ \citep{2012ApJ...753...48N} or $\sim\!2.46$ \citep{2012MNRAS.419.2195M} are inferred.

A striking feature called the "lithium dip" (Li-dip), 
was first observed in
(Population I) main-sequence stars 
in the Hyades open cluster by
\cite{1965ApJ...141..610W}, and confirmed by later studies \citep{1986ApJ...302L..49B,2000A&A...354..216B,2016ApJ...830...49B}.
Li abundances show a significant drop in the temperature range
$6400-6850\,\mathrm{K}$. Within this narrow temperature range, the depletion in A(Li) 
can reach a factor of 100 relative to stars out of this region. On the warm
side of the Li-dip, the Li abundances increase sharply with increasing effective temperature. For $\teff$ larger than $\sim$ 6900\,K, the Li abundances seem to remain constant, compatible with the Galactic value \citep[i.e., the meteoritic
value; see][]{2009LanB...4B..712L}. However, few abundance determinations are
available for stars in this $\teff$ region, because the primary abundance diagnostic, the \ion{Li}{I} 670.8 nm resonance line, is weaker in warmer 
stars; excessive line broadening due to typically fast stellar rotation further complicates spectroscopic analyses.
On the cool side of the Li-dip, the Li abundances increase gradually with decreasing effective temperature until reaching a sort of plateau, which extends from $\sim$ 6400 to $\sim$ 6000\,K. Stars in this region 
are slightly depleted in lithium, however this depletion is
uniform, and is not nearly as severe as in the Li-dip stars. 

Since the first observations, the presence of the Li-dip has also been found in many older star clusters, such as NGC 752 and M67 \citep{1995ApJ...446..203B},
but not in the youngest open clusters \citep{1988ApJ...327..389B,
2011MNRAS.410.2526B} -- those with ages less than about 
$100\,\mathrm{Myr}$. This indicates
that the large lithium depletions that are now apparent in the 
Li-dip take place when the stars are on the
main-sequence, rather than being there from the star's birth,
or occuring when the star was on the pre-main-sequence phase.

In order to meet observational constraints such as the complicated Li abundance
behaviour observed in the main-sequence stars, several different non-standard
stellar evolution models have been proposed, which take into consideration
atomic diffusion \citep{1986ApJ...302..650M} and rotation-induced mixing
\citep{1992A&amp;A...265..115Z}. However, these models cannot accurately
account for the observed Li-dip.  More recent works successfully managed to
describe the Li-dip in young stellar cluster Hyades by also accounting for
internal gravity waves \citep{2000A&amp;A...354..943M, 2005EAS....17..167C}.
According to their models, one can describe the Li-dip feature by
characterising the stars into three groups based on temperature
\citep{2005EAS....17..167C}: those warmer than the Li-dip, those within the
Li-dip, and those cooler. 
Stars in the warm group have the shallowest convective envelopes,
making them nearly unaffected by diffusion and rotation-induced mixing. 
Li-dip stars experience rotational mixing as the convective envelope deepens, resulting in severe Li destruction. But for stars in the cool group, even though they have even deeper convective envelopes, internal gravity waves become activated and efficiently extract angular momentum from the interior; this counter-acts the rotational mixing, and limits the amount of lithium destruction. For these reasons, stars on either side of the 
Li-dip have mechanisms to prevent lithium destruction to  different extents. In particular, the warm stars may allow us to probe the primordial Li abundance.

The Li-dip phenomenon
has also been observed in unevolved field stars \citep[e.g.][]{1999A&amp;A...348..487R,2001A&amp;A...371..943C,
2004MNRAS.349..757L,2012ApJ...756...46R,2018A&amp;A...615A.151B,2018A&amp;A...614A..55A}. These earlier studies have typical sample sizes of 200-2000 field stars in total, thus spanning only
very limited ranges in stellar properties and containing
very few stars on the warm side of the Li-dip. To study the 
lithium evolution comprehensively, a large sample of stars with
homogeneous measurements is needed.  

The aim of the present paper is to investigate the behaviour of 
lithium among late-type field stars including main-sequence,
turn-off and early sub-giant phase. Using data from the
Galactic Archeology with HERMES (GALAH) survey \citep{2015MNRAS.449.2604D},
we present the largest sample of lithium abundances so far.  The data span benefit from a homogeneous determination of stellar parameters and lithium abundances, and span a wide range of metallicities. These two aspects of our study allow us to draw fresh insights into the lithium puzzle.

\section{Observations and analysis} \label{sect:data} 

\begin{figure*} 
    \includegraphics[width=\textwidth,height=7cm]{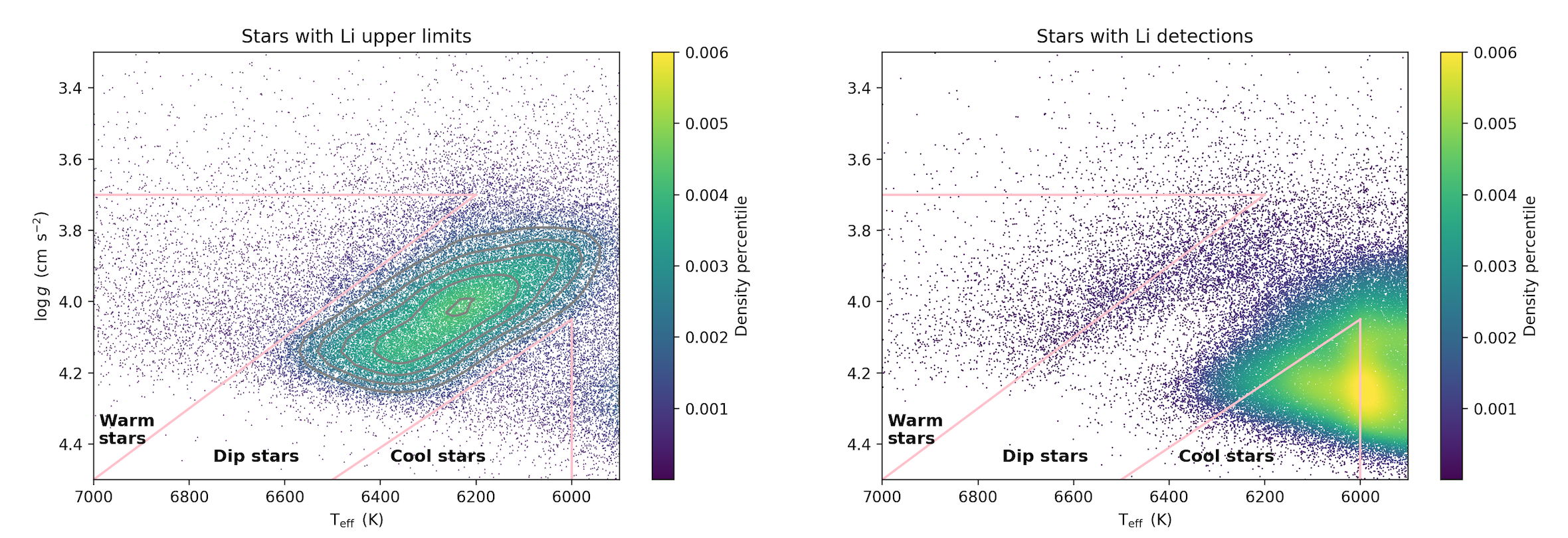}
    \caption{\textbf{Loci of the sample stars with 
    A(Li) upper limits and 
    lithium detections in the $\teff$-$\lgg$ panel, respectively.} \textbfit{Left
    panel}: Loci of stars with lithium upper limits
    (62\,945 stars).
    The colour bar represents
    density distribution of stars, with bright colours implying high density.
    The contours that are overplotted, represent the density percentile of
    lithium upper limit stars. We define the approximate boundaries of the
    Li-dip region by using linear fits in the left and right edges of the
    outermost contour. The boxes that are located inside the pink triangles are
    classified as 
    the warm group and cool group of stars, respectively. Stars that lie
    outside these three regions are removed from the comparison in
    \fig{fig:AlivsFeh}. \textbfit{Right panel}: Loci of stars with 
    lithium detections
    (59\,117 stars)}. \label{fig:HR_diagram} \end{figure*}

We observed over $650\,000$ FGK field stars in the solar neighbourhood as part of the GALAH \citep[]{2015MNRAS.449.2604D},  K2-HERMES
\citep{2019arXiv190412444S} and TESS-HERMES \citep{Sharma2017} spectroscopic
surveys. The spectral resolving power of the surveys $\mathrm{R} =
\frac{\lambda}{\Delta \lambda} \approx 28\,000$ is sufficiently matched to the stellar absorption lines under study. To avoid stars with large
convection-driven 
lithium depletion, we mainly target the dwarf and sub-giant stars
with $\teff$ ranging from 5900 to $7000\,\mathrm{K}$ covering a large range of
metallicity from $\feh=+0.5$ to $\feh=-3.0$, which includes the Spite plateau
at the metal-poor end. After excluding spectroscopically resolved binaries, and
observations with fitting inaccuracies, low signal-to-noise ratio, strong
emission lines or reduction issues, we obtain a set of 62\,945 stars with 
lithium detections and a separate set of 59\,117 stars with upper limits on the Li abundance.

The stellar parameters $\teff$, $\feh$, projected rotational velocities
$\vsini$, and line-of-sight radial velocity were determined simultaneously in a
homogeneous way by fitting observed neutral and ionized lines of Sc, Ti, and Fe
lines that are unblended and for which reliable atomic data are available, as
well as the $\teff$-sensitive $\halpha$~and $\hbeta$ lines, using the GALAH
analysis pipeline \citep{2018MNRAS.478.4513B}. Surface gravities were
constrained consistently and simultaneously by the fundamental relation between
the absolute magnitude, mass and $\teff$ \citep{2019A&amp;A...624A..19B}.
Stellar masses and ages were estimated by a Bayesian implementation of
isochrone fitting \citep{2018MNRAS.477.2966L}.  Having obtained the optimal
stellar parameters, Li abundances were then derived using non-local
thermodynamic equilibrium spectral synthesis \citep{2018MNRAS.481.2666G} for
the \ion{Li}{I} 670.8\,nm resonance line.  The non-LTE departure coefficients
come from the model described in \citep{2009A&amp;A...503..541L}.

In this work, we consider 
lithium to be detected when the line depression ($i.e.,
\mathrm{D} \equiv1-\frac{\digamma_{\lambda}}{\digamma_{c}}$) is deeper than
$1.5\sigma$ of the flux error within the line mask, and at least $3\%$ below
the normalised continuum flux.  In all other cases, the measurement is
considered as an upper limit.  We estimate upper limits on the Li abundance
based on linear interpolation in four dimensions, using a large matrix that
connects line strength with Li abundance, effective temperature, surface
gravity, and rotational velocity.

\section{Results} \label{sect:results}

\fig{fig:HR_diagram} shows the locations of the sample stars with 
lithium detections
and upper limits in the Kiel diagram, respectively.  Comparing the two panels, a clear gap is seen in the distribution of stars for which 
lithium could be detected (Fig.~1b), whereas a significant overdensity of stars for which only upper limits on the Li abundance could be obtained is seen in the same region (Fig.~1a).  Most of the stars with upper limits are concentrated in this
diagonal region between $\teff \sim 6300$ to $6600\,\mathrm{K}$, with surface gravity ($\lgg$) ranging from 3.8 to $4.3\,\dex$.  We interpret this as the Li-dip region: these stars have experienced severe lithium depletion and are now evolving towards to the sub-giant branch. 

To characterise the  Li-dip region in the Kiel diagram, we first narrow down the $\lgg$ range of our sample ($3.7-4.5\,\dex$) to reduce the the evolutionary effects on lithium due to post-main-sequence stars. Our sample now consists of upper main-sequence stars, turn-off stars and early sub-giants. Moreover, we remove all the stars with $\teff$ less than $6000\,\mathrm{K}$, as those cooler stars undergo strong and rapid 
lithium depletion, due to their larger convective envelopes
\citep[e.g.][]{2018A&amp;A...615A.151B}.  The density contours are then
overplotted on the distribution of A(Li) upper limits. We define the approximate boundaries of the  Li-dip region by using linear fits in the left and right edges of the outermost contour.  We delineate the left and right boundaries of the Li-dip region as the warm group and cool
group of stars, respectively.


\begin{figure*} \includegraphics[height=6.5cm,width=\textwidth]{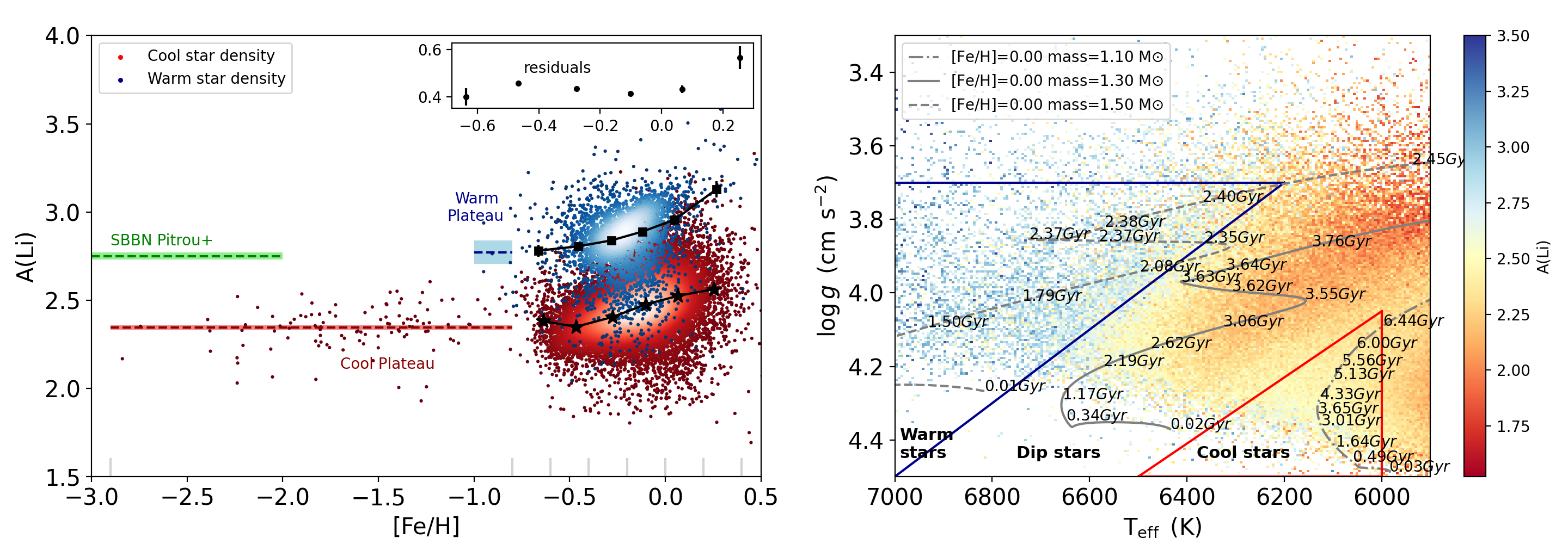}
    \caption{\textbf{Lithium trends as a function of metallicity 
    as observed for the 
    warm and cool groups of stars.} \textbfit{Left panel}:
    Observational data of 
    lithium from the warm and cool groups as a function of
    $\feh$. Only detections are retained in each group and plotted as
    colour-coded density. Black squares and asterisks represent the median Li abundances of the metallicity bins in the warm and cool groups of stars,
    respectively. The corresponding metallicity bin sizes are indicated with grey lines above the bottom axis.
    The error bars represent the uncertainties of the mean, most
    of them are too small to be shown. The residual of the mean Li abundances
    between the two groups in the corresponding metallicity range is shown as
    an inset in the upper right corner. The mean values of stars in the most
    metal-poor bin that stretches between $-2.9<\rm[Fe/H]<-0.8$ are instead
    marked with blue and red dashed lines for the 
    warm and cool groups of stars
    respectively, and marked as ``Warm Plateau" (6 stars) and ``Cool Plateau" (108 stars).  The
    corresponding shaded areas represent the standard error of the mean. The
    primordial Li abundance from the standard BBN
    \protect\cite{2018PhR...754....1P} is shown with a green dashed line, which is in close agreement with the Warm Plateau within the errors.
    \textbfit{Right panel}: All the stars with 
    lithium detections and upper limits
    are colour-coded by A(Li) in the $\teff$-$\lgg$ panel. The corresponding
    location of the warm and cool groups of stars are shown in the blue and red boxes, respectively. The colour gradient of Li abundance clearly shows the lithium differences between the three regions. Evolutionary tracks of different masses in the solar metallicity are overplotted.} \label{fig:AlivsFeh}
\end{figure*}

The left panel of \fig{fig:AlivsFeh} shows lithium trends as a function of metallicity, for the warm and cool groups of stars.
At low metallicity ($\feh \lesssim -1\,\dex$), the stars in the cool group have ages in excess of
$11\,\mathrm{Gyr}$ and reveal the Spite Plateau, showing low and near-constant
Li abundances.  In contrast, there are no old, metal-poor warm stars in our
sample: such stars have higher masses and have evolved off the main-sequence
after such a long period. Thus the stars in the warm group only appear above $\rm[Fe/H]\gtrsim -1.0$. Up to $-1.0\lesssim \feh \lesssim-0.5$, the warm group having 117 stars shows a similar constant lithium plateau, but elevated by almost three times that of the cool group ($0.4\,\dex$).
We measure A(Li) $=2.69\pm0.06$.
Remarkably, this plateau
is largely consistent with the predictions of BBN (A(Li) $=2.75\pm0.02$).
One interpretation of this, is that 
both Galactic enrichment and stellar destruction have been
insignificant in this population of stars;
in other words, that these stars in fact may
preserve the primordial Li produced in the early Universe.

At higher metallicities $\feh \gtrsim -0.5$, 
both the warm and cool groups of stars show an increasing
trend in A(Li). This is probably caused by
Galactic enrichment \citep{2017A&amp;A...606A.132P}.
It is interested to note that 
even in this metallicity regime,
the difference in the average Li abundances between the warm and
cool groups is still $0.4\,\dex$, and remains so up to
solar metallicity.

The right panel of \fig{fig:AlivsFeh} shows the corresponding location of warm and cool group stars in the Kiel diagram with color-coded Li abundances. There is a clear gradient in the Li abundances across the Kiel diagram, delineating the Li-dip from the warm and cool groups of stars. 
It shows that stars in the cool group are systematically
more depleted in lithium than those in the warm group. 
Since most of our stars are
centered around the solar metallicity, we overplotted the evolutionary tracks
of different masses in solar metallicity on the distribution of our targets.
The theoretical models support our
speculation that we have captured the evolutionary track of 
Li-dip stars in our observations.

At a given $\feh$, there is a difference in mean age between the
warm group (young) and cool group (old), because of the sample selection
method.  The age difference is largest at low metallicity (up to
$7\,\mathrm{Gyr}$) and steadily decreases to become insignificant at the
highest metallicities. The lower Li abundances of the cool group should thus
be interpreted as a combined effect of their lower effective temperatures and older ages, making depletion more efficient and giving it longer time to act.
\cite{2005A&A...442..615S} carried out a detailed investigation of
lithium depletion time scales in late-type stars at this temperature with metallicities $-0.2\lesssim\feh\lesssim+0.2$. They found that main-sequence lithium depletion is not a continuous process and becomes ineffective beyond an age of 1-2\,Gyr for the majority of stars, leading to a Li plateau at older ages. The total amount of lithium depletion during the main-sequence lifetime is around 0.5\,dex. 

The age difference between the warm and cool groups may lead one to speculate that Galactic chemical evolution has elevated the initial Li abundances in the warm group compared to in the cool group.
However, our data suggests that such enrichment scales with increasing
metallicity and only becomes noticeable at $\feh\gtrsim-0.5$ where 
\fig{fig:AlivsFeh} above shows the stars in both the warm and cool groups clearly rise off their respective plateaus.  Moreover, recent observations of Li abundances in the low metallicity gas
($\feh\approx-0.5$) of the Small Magellanic Cloud 
\citep{2012Natur.489..121H}
and in warmer stars in the open cluster NGC 2243 \citep{2013A&amp;A...552A.136F}
($\feh=\,-0.52$; an estimated age of $4.3\,\mathrm{Gyr}$) are in good agreement
with mean A(Li) measurements in the warm plateau stars. 
The value of Li abundance obtained from these observations are all comparable to the primordial Li abundance predicted by BBN.

\section{Conclusions} \label{sect:implications} 
To reduce the complex behaviour
of lithium in the field stars, for the first time we draw a comparison between the warm and cool groups of stars, 
which are located on the warm and cool side of the Li-dip, respectively. Here we find that Li abundances in the two groups show a
similar pattern as a function of $\feh$, however, stars in the cool group are more depleted in lithium than those in the warm group by a factor of three. This difference is determined from more than 100\,000 stars that have a wide range of stellar properties and chemistry. The implications we can obtain from this result are as follows.

\begin{itemize} \item We find that at $-1.0\lesssim\feh \lesssim-0.5$,
            the average Li abundance of stars warmer than 
            the Li-dip shows an elevated lithium plateau, the value of which
            is consistent with the primordial Li abundance as predicted by BBN within the errors.
            Since Galactic production of
            lithium has not yet contributed significantly to the cooler Spite plateau in this metallicity regime, we suggest that the Li abundance measured in the warm group is indicative of insigificant Li depletion as well as insignificant Li enrichment in these stars. This interpretation would explain why the Li abundances closely resemble the BBN primordial value, and is consistent with what has been reported for the Small Magellanic Cloud \citep{2012Natur.489..121H} and in the open cluster NGC 2243 \citep{2013A&amp;A...552A.136F}. Regardless of the possibility that minor depletion and enrichment may have cancelled each other out in this group of stars, the values provide a valuable constraint on both cosmological models of the early universe and stellar evolution models.

        \item We infer that, 
            at a given metallicity, 
            the three different regimes (warm, 
            Li-dip, cool) follow
            different lithium depletion mechanisms.
            For the stars in the cool group, the
            depletion is not strongly dependent of metallicity;
        instead it is primarily governed by a star's main-sequence temperature and age. How much lithium has
        been depleted is a combination of temperature and stellar age, causing a near-constant offset with respect to the warm group.

\item We identify $\feh \approx -0.5$ as the turning point where the Li
    abundances break the plateau and Galactic 
    lithium production becomes significant.
    We see this in both the warm and the cool groups of stars.
    Given that the Spite plateau stars (of the cool group) have already
    experienced a large depletion of lithium and therefore do not reflect
        the true primordial value, we recommend that modellers apply the BBN-predicted Li abundance, instead of the Spite plateau
\citep{1982A&amp;A...115..357S} Li abundance, as the initial value in chemical evolution models \citep{2012A&amp;A...542A..67P,2017A&amp;A...606A.132P}.
\end{itemize}

\section*{Acknowledgements}
\label{acknowledgements}

XDG, KL, AMA and SB acknowledge funds from the Alexander von Humboldt Foundation in the framework of the Sofja Kovalevskaja Award endowed by the Federal Ministry of Education and Research. 
KL also acknowledges funds from the Swedish Research Council (VR 2015-004153) and Marie Sk{\l}odowska Curie Actions (cofund project INCA 600398), and AMA also acknowledges support from the Swedish Research Council (VR 2016-03765),  and the project grant `The New Milky Way' (KAW 2013.0052) from the Knut and Alice Wallenberg Foundation.
TZ acknowledges financial support of the Slovenian Research Agency (research core funding No. P1-0188). SLM and JDS
acknowledges the support of the Australian Research Council through Discovery Project grant DP180101791. Parts of this research were conducted by the Australian Research Council Centre of Excellence for All Sky Astrophysics in 3 Dimensions (ASTRO 3D), through project number CE170100013. YST is grateful to be supported by the NASA Hubble Fellowship grant HST-HF2-51425 awarded by the
Space Telescope Science Institute. SWC acknowledges federal funding from the Australian Research Council though the Future Fellowship grant entitled `Where are the Convective Boundaries in Stars?' (FT160100046).
GT acknowledges support by the project grant `The New Milky Way' from the Knut and Alice Wallenberg foundation and by the grant 2016-03412 from the Swedish Research Council.
This work is also based
on data acquired from the Anglo-Australian Telescope. We acknowledge the
traditional owners of the land on which the AAT stands, the Gamilaraay people,
and pay our respects to elders past and present.

\bibliographystyle{mnras}
\bibliography{bib_file}

\label{lastpage}
\end{document}